\documentclass[superscriptaddress,amsmath,amssymb,twocolumn,floatfix]{revtex4-2}

\usepackage[noindentafter]{titlesec}
\usepackage{titlesec}
\titleformat{name=\section}
  {\normalfont\large\bfseries\MakeUppercase}{\MakeUppercase{\thesection}}{0pt}{}
\titleformat{name=\subsection}
  {\normalfont\bfseries}{\thesection}{0pt}{}
\titlespacing{\section}{0cm}{0.7cm}{0.01cm}
\titlespacing{\subsection}{0cm}{0.45cm}{0cm}

\usepackage[utf8]{inputenc}
\usepackage{listings}
\usepackage{footmisc}
\usepackage{braket}
\usepackage{graphicx}
\usepackage[caption=false]{subfig}
\usepackage[colorlinks=True,linkcolor=red,citecolor=blue,urlcolor=blue]{hyperref}

\usepackage{blkarray}
\usepackage{array}
\usepackage[dvipsnames]{xcolor}
\usepackage[normalem]{ulem}

\usepackage{xspace}
\newcommand{\rucl}{$\text{RuCl}_\text{3}$\xspace}
\newcommand{\rubr}{$\text{RuBr}_\text{3}$\xspace}
\newcommand{\rui}{$\text{RuI}_\text{3}$\xspace}
\newcommand{\rux}{$\text{RuX}_\text{3}$\xspace}

\usepackage{tabularx}
\usepackage{array}   
\newcolumntype{L}{>{$}l<{$}} 
\newcolumntype{R}{>{$}r<{$}} 
\newcolumntype{C}{>{$}c<{$}}

\usepackage{float}

\usepackage[capitalise]{cleveref}
\usepackage{placeins}

\date{\today}

\begin{document}
	\title{%
	Electronic and magnetic properties of the RuX$_3$ (X=Cl, Br, I) family:\\ Two siblings  --- and a cousin?
}
\author{David A. S. Kaib}
\email{kaib@itp.uni-frankfurt.de}
\affiliation{Institut f\"ur Theoretische Physik, Goethe-Universit\"at Frankfurt, 60438 Frankfurt am Main, Germany}
\author{Kira Riedl}
\email{riedl@itp.uni-frankfurt.de}
\affiliation{Institut f\"ur Theoretische Physik, Goethe-Universit\"at, 60438 Frankfurt am Main, Germany}
\author{Aleksandar Razpopov}
\affiliation{Institut f\"ur Theoretische Physik, Goethe-Universit\"at, 60438 Frankfurt am Main, Germany}
\author{Ying Li}
\affiliation{Department of Applied Physics and MOE Key Laboratory for Nonequilibrium Synthesis and Modulation of Condensed Matter, School of Physics, Xi'an Jiaotong University, Xi'an 710049, China}
\author{Steffen Backes}
\affiliation{CPHT, CNRS, Ecole Polytechnique, Institut Polytechnique de Paris, Route de Saclay, 91128 Palaiseau, France}
\author{Igor I. Mazin}
\affiliation{Department of Physics and Astronomy and Quantum Science and Engineering Center, George Mason University, Fairfax,
Virginia 22030, United States}
\author{Roser Valent\'i}
\email{valenti@itp.uni-frankfurt.de}
\affiliation{Institut f\"ur Theoretische Physik, Goethe-Universit\"at, 60438 Frankfurt am Main, Germany}

	\date{\today}
	
	\begin{abstract}
 Motivated by reports of metallic behavior in
		the recently synthesized RuI$_3$, in contrast to
		the Mott-insulating nature of the actively discussed
		$\alpha$-RuCl$_3$, as well as  RuBr$_3$, we 
		present a detailed comparative analysis of the electronic and magnetic
		properties of this family of trihalides. Using a combination of first-%
		principles calculations and effective-model considerations,
		we conclude that
		RuI$_3$, similarly to the other two members,
		is most probably on the verge of a Mott insulator, but with much smaller magnetic moments
		and a strong magnetic frustration. 
		We predict the ideal pristine crystal of \rui to have a nearly vanishing conventional nearest-neighbor Heisenberg interaction and
		to be a quantum spin liquid candidate of possibly different kind than the Kitaev spin liquid.
		In order to understand the apparent contradiction to
		the reported resistivity $\rho$,
		we analyze the experimental evidence for all three compounds and propose
		a scenario for the observed metallicity in existing samples of RuI$_3$.
		Furthermore, for the Mott insulator \rubr we obtain a magnetic Hamiltonian of a similar form to that
		in the much discussed $\alpha$-RuCl$_3$  and show that this
		Hamiltonian is in agreement with experimental evidence in \rubr. %

	\end{abstract}
    
\maketitle

\section{Introduction}
\begin{figure*}
    \centering
    \includegraphics[width=\textwidth]{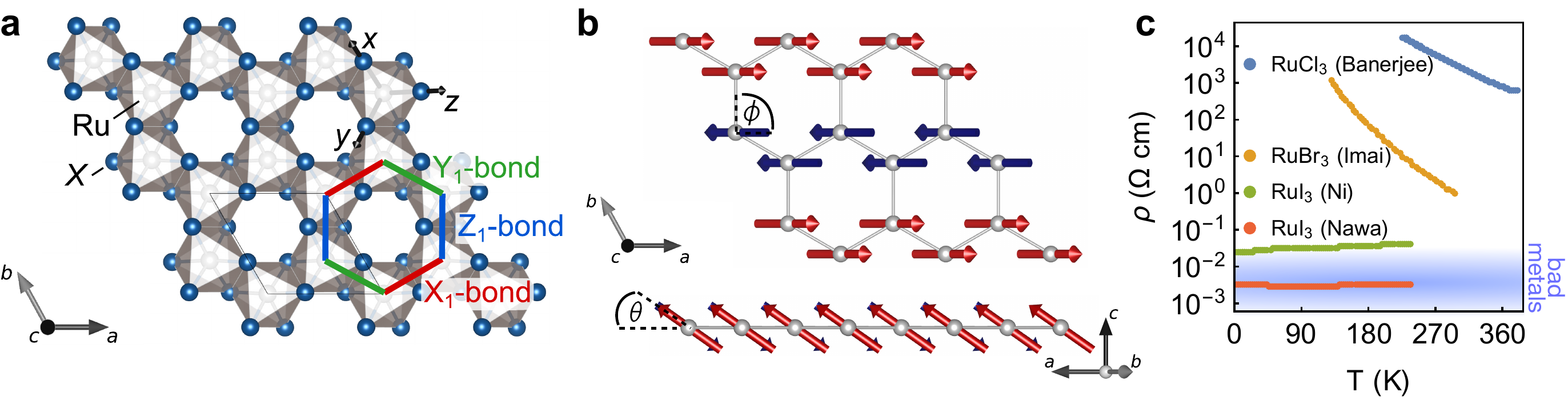}
    \caption{\textbf{\rux (X=Cl, Br, I) crystal structure, magnetic structure and resistivity.} 
   \textbf{a}~Honeycomb layer in the \rux (X=Cl, Br, I) trihalides with bond definitions,
    cubic axes $(xyz)$ and crystallographic axes $(abc)$ in the R$\bar{3}$ structure, \textbf{b}~Zigzag magnetic order in a honeycomb layer from two perspectives, with definitions of in-plane-angle $\phi$ and out-of-plane-angle $\theta$. \textbf{c}~Comparison of experimental dc resistivities
    as a function of temperature. Data was extracted from plots in the following references and labelled by respective first-author names: 
    \rucl~(Banerjee~\cite{banerjee2017NeutronScatteringProximate}), %
    \rubr~(Imai~\cite{imai2021magnetism}), 
\rui~(Ni~\cite{danrui2021honeycomb}, Nawa~\cite{nawa2021StronglyElectroncorrelatedSemimetal}). The shaded background depicts a typical range of resistivity for bad metals \cite{jaramillo2014origins}.
    \label{fig:structure}}
\end{figure*}

\rui and \rubr are recent additions to the \rux family (X= Cl, Br, I)
of layered Ru-based trihalides (Fig.~\ref{fig:structure}a). The first member, $\alpha$-RuCl$_3$ (in the following `RuCl$_3$') has attracted considerable attention in recent years  as a candidate material for the Kitaev honeycomb model~\cite{kitaev2006anyons}. 
\rucl is a spin-orbit assisted Mott insulator \cite{jackeli2009mott,plumb2014honeycomb,johnson2015,zhou2016photoemission} whose magnetic low-energy degrees of freedom can be described in terms of $j_\mathrm{eff}=1/2$ moments that interact through strongly anisotropic exchange~\cite{jackeli2009mott,rau2014generic,kim2016crystal,winter2016challenges}. While the material enters a so-called zigzag antiferromagnetic order (\cref{fig:structure}b) at low temperatures $\ensuremath{T_\text{N}}\approx 7\,$K \cite{johnson2015,cao2016,banerjee2017NeutronScatteringProximate}, various
experiments at finite temperature~\cite{sandilands2015scattering,nasu2016fermionic,do2017majorana,widmann2019thermodynamic} or at finite magnetic field~\cite{johnson2015,sears2017phase,banerjee2018excitations,kasahara2018majorana,hentrich2018unusual} have been interpreted as hallmarks of Kitaev
physics, a subject which is presently under intensive 
debate~\cite{winter2017breakdown,hentrich2020high,sahasrabudhe2020high,chern2021sign,czajka2021oscillations,lefranccois2021evidence}. 

Recently, a sister compound with a heavier halogen, $\text{X}=\mathrm{Br}$, was synthesized \cite{imai2021magnetism}. Analogous to \rucl, it is insulating and shows zigzag magnetic order, albeit with higher Néel temperature $T_{\mathrm N}=34$\,K \cite{imai2021magnetism}. In contrast to \rucl, the authors of Ref.~\onlinecite{imai2021magnetism} reported a Weiss constant with dominant {antiferromagnetic} interactions and a direction of the zigzag ordered moment
different from \rucl, and argued that this deviation suggests a closer proximity to the pure Kitaev model.

To complete the \rux family, two independent groups have now synthesized \rui  
with the even heavier halogen iodine \cite{nawa2021StronglyElectroncorrelatedSemimetal,danrui2021honeycomb}. 
 In contrast to the two `sibling' compounds, a quasi-metallic behavior was observed in \rui, questioning the description in terms
of localized $j_\mathrm{eff}=1/2$ moments.
Even though the dc resistivities measured in \rui are orders of magnitude smaller than those of 
\rucl or \rubr, the reported values of $10^{-3}$ to $10^{-2}\ \Omega\,\mathrm{cm}$
\cite{nawa2021StronglyElectroncorrelatedSemimetal} are uncharacteristically large for metals or even
typical {bad metals} \cite{jaramillo2014origins}, and practically temperature-independent. 
While neither of the groups found clear signatures of magnetic ordering \cite{nawa2021StronglyElectroncorrelatedSemimetal,danrui2021honeycomb}, they reported different behaviors of the magnetic susceptibility, which is either found to be temperature-independent \cite{nawa2021StronglyElectroncorrelatedSemimetal}, or with a strong upturn at low temperatures \cite{danrui2021honeycomb}, suggesting that sample quality plays a crucial role. 

In order to understand the apparently distinct behavior of this family of trihalide
materials, in this work
we analyze the available experimental data and perform a detailed
comparative study of the electronic and magnetic properties of the systems via
first-principles calculations and extracted low-energy models. 
We find that:
 (i)~The behavior
of \rui is not
that far from \rucl and \rubr and the variations across the series
are more quantitative than qualitative. 
(ii)~Pristine samples of \rui should be insulating with highly anisotropic magnetic exchange 
and nearly vanishing conventional Heisenberg interaction. %
We argue that the reported metallic behavior in \rui could have its origin in sample quality.
(iii)~The magnetism in the Mott insulator
\rubr has predominantly ferromagnetic interactions, in contrast to what is suggested by the Curie-Weiss analysis of Ref.~\cite{imai2021magnetism}. We show that such interactions are consistent with experiment when taking into account spin-orbit coupling effects in the Curie-Weiss behavior. 

Our study derives model parameters and magnetic Hamiltonians for the whole \rux family from \textit{ab-initio}, that will be useful for future
theoretical studies of these systems. %
In contrast to the usual model derivations that only include local spin-orbit coupling (SOC) on the magnetic ion \cite{jackeli2009mott,rau2014generic,winter2016challenges}, our approach includes all SOC effects in the crystal. In fact, we show that SOC from the ligands leads to significant deviations from the Ruthenium-only SOC picture in the case of \rubr and \rui.

\section{Results and discussion}

\subsection{Comparative analysis of experiments}
In the following, we analyze the reported electrical resistivity, specific heat and magnetic susceptibility 
 data for \rux\ (X=Cl, Br, I)~\cite{banerjee2017NeutronScatteringProximate,little2017AntiferromagneticResonanceTerahertz,imai2021magnetism,danrui2021honeycomb,nawa2021StronglyElectroncorrelatedSemimetal}.

In \cref{fig:structure}c we summarize the 
temperature dependence of the experimental resistivity data \cite{banerjee2017NeutronScatteringProximate,danrui2021honeycomb,imai2021magnetism,nawa2021StronglyElectroncorrelatedSemimetal} in all three compounds.
In \rui, the resistivity  has a weak~\cite{danrui2021honeycomb} or almost no~\cite{nawa2021StronglyElectroncorrelatedSemimetal} temperature dependence
(\cref{fig:structure}c).
Traditionally,
 metals are classified as materials where the resistivity $\rho$ increases with temperature, 
 distinguishing {conventional metals} ($e.g.$, Cu) as those where in clean samples
 at temperatures roughly 300 to 600\,K, $\rho \sim 10^{-6}\ \Omega\cdot \mathrm{cm}$ to $\sim 10^{-5}\ \Omega\cdot\mathrm{cm}$, and {bad metals} as those
 with resistivities of
$\sim 1-10$\ m$\Omega\cdot$cm %
 This range is shown as a background shading in \cref{fig:structure}c.
  The reported
 resistivities for \rui 
 ($\rho\sim 40$\ m$\Omega\cdot$cm~\cite{danrui2021honeycomb}
 and $\rho\sim 4$\ m$\Omega\cdot$cm~\cite{nawa2021StronglyElectroncorrelatedSemimetal})
 are high even for bad metals, surpassing the Ioffe-Regel limit by more than an order of magnitude. Even more relevant, the {lower-resistivity} set of data~\cite{nawa2021StronglyElectroncorrelatedSemimetal} shows no discernible temperature dependence
 at all, while the data in Ref.\ \onlinecite{danrui2021honeycomb} show a very weak positive 
 derivative $d\rho/dT$, but the absolute value is above anything traditionally considered 
 metallic.

Seemingly, as also pointed out in Ref.~\onlinecite{danrui2021honeycomb}, 
electron transport in existing \rui samples may be contaminated by grain boundaries. One possibility to interpret the measurements is that the pristine material is metallic, but insulating grain boundaries prevent percolation. Then, the
in-grain resistivity can be neglected and what is measured is the resistivity of the insulating
grain boundaries. In that case, however, thermal activation of carriers in the boundaries should give a positive temperature gradient of the resistivity, which is not observed. The opposite scenario is that of an insulating behavior in the bulk and (possibly bad) metallic one between the grains. %
In that case, the large resistivity reflects the small relative volume of metallic boundaries, where the transport is dominated by the residual resistivity. This scenario is compatible with
the observations. Morphology of the grain boundaries can vary wildly
  depending on the growth conditions, including but not limited to vacancies,
  twins, dislocation and plain chemical dirt. Grain boundaries in
  semiconductors are often observed to be metallic. Apart from grain boundaries contaminating resistivity measurements, disorder (in form of vacancies, stacking faults, etc.\ \cite{danrui2021honeycomb,nawa2021StronglyElectroncorrelatedSemimetal}) could promote the bulk metallic phase over the Mott-insulating one, 
as has been shown for example for the Mott insulator $\kappa$-(BEDT-TTF)$_2$Cu[N(CN)$_2$]Cl
 \cite{gati2018effects}. Indeed, in our first-principles calculations discussed below, we find the ideal \rui\ to already be quite close to a Mott-metal transition.

Turning to the sibling compounds \rucl and \rubr, the resistivity (\cref{fig:structure}c) decreases with temperature, as expected for Mott insulators,
and both systems show an approximate exponential activation gap behavior, $E_{g,\mathrm{eff}}(T)=-k_\mathrm{B} T^2 (\mathrm d \operatorname{ln}\rho/\mathrm dT)$,
although with a significant blue-shift of the gap with increasing temperatures. %

Considering specific heat data in the compounds, 
the specific heat for \rucl displays a well-defined peak at $\ensuremath{T_\text{N}}\approx 7$\,K denoting the onset of the zigzag order, while the onset of long-range magnetic order in \rubr is
observed by a kink at $\ensuremath{T_\text{N}}=34$\,K \cite{imai2021magnetism}. 
None of this is observed for \rui~\cite{danrui2021honeycomb,nawa2021StronglyElectroncorrelatedSemimetal}.
In \cref{tab:specific_heat} we summarize specific heat parameters reported experimentally \cite{tanaka2020ThermodynamicEvidenceFieldangle,imai2021magnetism,danrui2021honeycomb,nawa2021StronglyElectroncorrelatedSemimetal}, where $\gamma$ ($\beta$) is the $T$-linear ($T^3$) contribution to $C(T)$.

In \rui, the $T$-linear contribution, even though contaminated by 
an extrinsic raise at small temperatures in Ref.~\cite{danrui2021honeycomb} attributed to the nuclear quadrupole moment of Ru, yields  $\gamma\sim 15-30$\,mJ$\cdot$K$^{-2}\cdot$mol$^{-1}$ \cite{danrui2021honeycomb,nawa2021StronglyElectroncorrelatedSemimetal}.  From our electronic structure calculations of \rui shown below, 
we find
that the unrenormalized metallic (\textit{i.e.}, nonmagnetic, not $U$-corrected)
density of states corresponds to $\gamma_0\approx 3$\,mJ$\cdot$K$^{-3}\cdot$mol$^{-1}$, suggesting a mass renormalization 
(if this $\gamma$ is intrinsic) of a factor of 7--12.  In the scenario where the metallic
grain boundaries take up a 
sizeable fraction of the sample volume, this renormalization 
shall be even stronger, encroaching into the heavy 
fermions domain. This suggests that the origin of the anomalously
large residual heat capacity may  not be related to intrinsic
metallicity. 
It is worth noting that the $T^3$ term $\beta$ of $C(T)$, on the other hand, 
is rather reasonable for the three systems and scales roughly as the harmonic average $M_{\text{RuX}_3}$
of the atomic masses (last column in \cref{tab:specific_heat}). %

\begin{table}
\centering
\begin{ruledtabular}
\begin{tabular}{l|c|c|c|cc}
     Sample & $\gamma$ & $\beta$ & $T_\text{D}$ & 
     $T_\text{D} M_{\text{RuX}_3}^{1/2}$
     \\
     \hline
     \rucl* (Tanaka \cite{tanaka2020ThermodynamicEvidenceFieldangle}) & & 1.22 & 185 & 1 \\
     \rubr (Imai \cite{imai2021magnetism})   &  & 1.93 & 159 & 1.21 \\
     \rui (Ni \cite{danrui2021honeycomb})  & 29.3 & 4.72 & 118 & 1.07\\
     \rui (Nawa \cite{nawa2021StronglyElectroncorrelatedSemimetal}) & 17.7 & 3.66 & 129 & 1.17
\end{tabular}
\end{ruledtabular}
\caption{\textbf{Overview of reported specific heat parameters.} $\gamma$ ($\beta$) is the coefficient of the $T$-linear ($T^3$) contribution and given in units of mJ\,K$^{-2}$mol$^{-1}$ (mJ\,K$^{-4}$mol$^{-1}$). Debye temperature $T_{\text D}=\left(\frac{12\pi^4 N R}{5\beta}\right)^{1/3}$ %
is given in Kelvin and $T_\mathrm{D} M_{\text{RuX}_3}^{1/2}$ as a ratio to the value for \rucl\ (first row), where $M_{\text{RuX}_3}$ is the harmonic average of the RuX$_3$ mass. (*)~Note that the values given for \rucl correspond to the asymptotic field-polarized limit extracted by Tanaka {\it et al.\ }\cite{tanaka2020ThermodynamicEvidenceFieldangle}, as otherwise at zero field the low-temperature specific heat behavior is dominated by vicinity to the Néel temperature of \rucl, causing large magnetic contributions to $\beta$. 
}
    \label{tab:specific_heat}
\end{table}

\begin{figure}
    \centering
    \includegraphics[width=\linewidth]{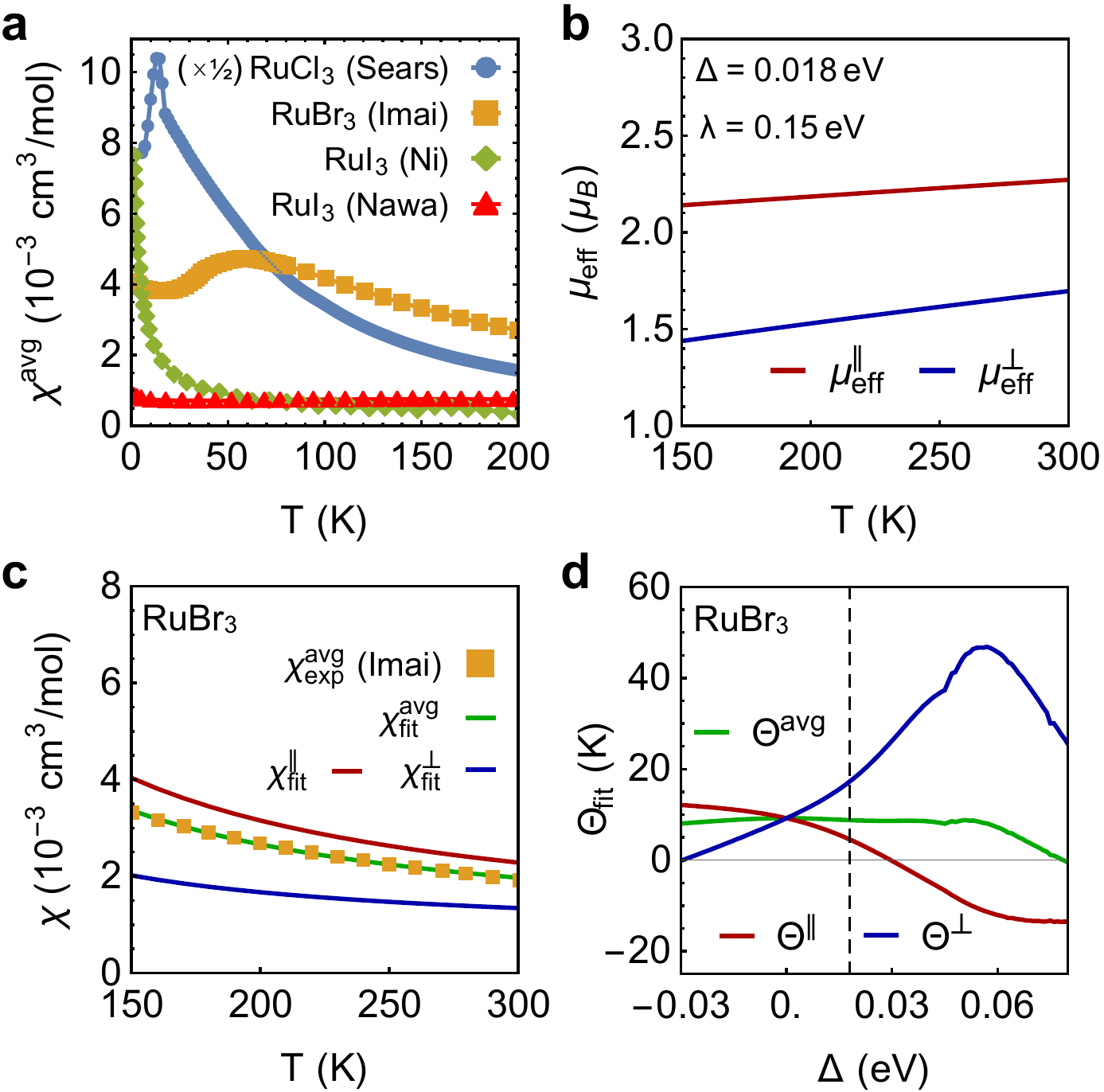}
    \caption{\textbf{Magnetic susceptibility and modified Curie-Weiss fit.}  \textbf{a}~Experimental direction-averaged \rux susceptibility data, extracted from plots in the following references and labelled by respective first-author names: 
    \rucl (Sears) \cite{sears2015MagneticOrderRuCl}, 
    \rubr\ (Imai) \cite{imai2021magnetism}, 
\rui (Ni)  \cite{danrui2021honeycomb}, \rui\ (Nawa)  \cite{nawa2021StronglyElectroncorrelatedSemimetal}. Note that the \rucl\ curve is scaled by~$\frac12$. %
\textbf{b}~Calculated temperature-dependent effective moment $\mu_{\mathrm{eff}}(T)$ for $\Delta=0.018$\,eV and SOC $\lambda=0.15$\,eV. 
\textbf{c}~Modified Curie-Weiss fit of \rubr data, taking into account such $\mu_\mathrm{eff}(T)$. 
\textbf{d}~Dependence of best-fit Weiss constants on assumed $\Delta$. Vertical dashed line indicates $\Delta=0.018$\,eV. 
}
    \label{fig:susceptibility}
\end{figure}

We now turn our attention to magnetic susceptibility measurements.
\Cref{fig:susceptibility}a summarizes the powder-averaged measured magnetic susceptibilities $\chi(T)$
as reported in Refs.~\cite{sears2015MagneticOrderRuCl,imai2021magnetism,danrui2021honeycomb,nawa2021StronglyElectroncorrelatedSemimetal}. 
At low temperatures the \rucl data \cite{sears2015MagneticOrderRuCl} shows a clear signature of a transition to the ordered magnetic phase
at $7\,$K. For \rubr, the Néel transition $T_\mathrm{N}\approx 34\,$K is less apparent from the susceptibility, but the maximum in $\mathrm d\chi/\mathrm dT$ is consistent with the distinct transition seen in NMR relaxation measurements \cite{imai2021magnetism}.
The experimental report on powder samples of \rubr
utilized a standard Curie-Weiss (CW) fit, yielding an average Curie-Weiss temperature $\Theta^{\rm avg}_{\rm std}=-58$\,K \cite{imai2021magnetism}, indicating predominantly AFM interactions. However, as we have recently shown~\cite{li2021ModifiedCurieWeissLaw}, the Weiss constants obtained with such a standard CW fit may not anymore reflect
the intrinsic exchange couplings in the case of significant SOC %
  in the material, as it is the case for the Ru-based trihalides.  With SOC, %
 temperature-dependent van-Vleck contributions can arise, which can be effectively captured in a temperature-dependent magnetic moment $\mu_\mathrm{eff}(T,\Delta)$ \cite{li2021ModifiedCurieWeissLaw}, as shown for $\Delta=0.018$\,eV in \cref{fig:susceptibility}b,
 where $\Delta$ can be
 directly associated to the crystal field splitting resulting
from the distorted octahedral environment of Ru.
In fact, for the sister compound \rucl, a standard CW fit would lead to $\Theta^{\rm avg}_{\rm std}=-20$\,K, whereas 
an improved CW fit taking into account such \textit{van-Vleck}-like  contributions \cite{li2021ModifiedCurieWeissLaw}
provides 
CW constants  $\Theta^{\parallel} = +55\,$K for the magnetic field in the honeycomb plane and $\Theta^{\perp}= +33\,$K for the out-of-plane
field, revealing an average CW constant, $\Theta^{\rm avg} =\frac{2\Theta^{\parallel}+\Theta^{\perp}}{3}$, of $\approx 48$K. This indicates predominant ferromagnetic (FM) interactions, as they have become established for the magnetic Hamiltonian in  \rucl~\cite{winter2017models,laurell2020dynamical,sears2020ferromagnetic,suzuki2021proximate}.

Considering a similar strategy (see `Methods' section), we fit the average susceptibility $\chi^{\text{avg}}$ of \rubr \cite{imai2021magnetism}. 
However, since the crystal-field parameter $\Delta$ primarily controls the in-plane vs out-of-plane anisotropy, and for \rubr only powder-averaged data are available, we do not aim at extracting $\Delta$ by fitting. Instead, we first
fix $\Delta$ using our first-principles calculations, enforcing $\mu_{\rm eff}^{\parallel}/\mu_{\rm eff}^{\perp} (T=0K)$ $\propto$  $g_{\parallel}/g_{\perp}$, where $g_{\parallel}/g_{\perp}$ are taken
from quantum chemistry calculations (see \cref{fig:exchange_parameters}a, discussed below). This leads to $\Delta=0.018$\,eV. 
The best CW fit accounting for the implied $\mu_\mathrm{eff}(T,\Delta=0.018\,\mathrm{eV})$ (shown in \cref{fig:susceptibility}c) yields Weiss constants $\Theta^{\parallel}\approx 5$\,K, $\Theta^{\perp}\approx 17$\,K and $\Theta^{\mathrm{avg}}\approx 9$\,K, which are positive, indicating predominately ferromagnetic interactions for \rubr, as seen before in \rucl. 
In \cref{fig:susceptibility}d we further analyze how the best-fit Weiss constants evolve for other choices of $\Delta$. Indeed, for a wide range of reasonable $\Delta$ around the first-principles value 
(indicated by the dashed vertical line), the average Weiss constant $\Theta^{\rm avg}$ remains positive.

Importantly, for both materials, a standard residual `background' term has
to be included in the fitting, which in our case, depending on the material (\rubr or \rucl) ranges from $\sim -3.5\times 10^{-4}$  emu/mol to $1.5\times 10^{-4}$.
This is of the same order of magnitude as the corresponding
term in \rui ($\sim 3$ to $8\times 10^{-4}$\,emu/mol) \cite{nawa2021StronglyElectroncorrelatedSemimetal,danrui2021honeycomb}. 
Since in the former cases an intrinsic Pauli origin can be excluded, this observation
also casts doubts on 
a metallic interpretation of this term in \rui.
Actually, the two available susceptibility measurements on \rui display different behaviors, one nearly temperature-independent~\cite{nawa2021StronglyElectroncorrelatedSemimetal}, 
  and the other~\cite{danrui2021honeycomb} 
 showing a Curie-like rise at low temperatures,
 where a standard CW fit yields $\mu_{\rm eff}=0.53\mu_\mathrm{B}$ and $\Theta_{\rm CW}^{\rm avg}=-3$\,K ~\cite{danrui2021honeycomb}. 
These differences are consistent with our hypothesis that the measured samples consist of magnetic insulating grains
surrounded by metallic boundaries.  
Then, the samples with larger resistivity data~\cite{danrui2021honeycomb} hint to larger insulating grains, hence less metallic boundaries are present, leading to the low-temperature Curie-like upturn in the susceptibility, compared to the samples in Ref.~\onlinecite{nawa2021StronglyElectroncorrelatedSemimetal}.

\subsection{Electronic and magnetic calculations}
In the following we present a comparison of the electronic and magnetic properties 
of the trihalide \rux family obtained from a combination of density functional theory (DFT)
and exact diagonalization of \textit{ab-initio}-derived low-energy models.
Details of the calculations are given in the "Methods" section.

Past experience with first-principles calculations for
the Ru-based 
trihalides~\cite{johnson2015,kim2016crystal,kaib2021magnetoelastic,kim2021spin,zhang2021}
indicates that the
magnetic order and, to a considerably lesser extent, metallicity
is very fragile, with several closely competing different magnetic phases.
The ground states may vary depending on small changes in the
crystal structure, on the way in which strong
correlations are accounted for, and even on tiny details of the
computational protocol. With this in mind, it is imperative to
compare the calculated properties across the
series, using the exact same computational setup.

For the electronic structure calculations we consider  the experimentally reported C2/m \cite{johnson2015,cao2016} and R$\bar{3}$ \cite{park2016} structures for \rucl, and the 
suggested R$\bar{3}$ structures for \rubr \cite{imai2021magnetism} and \rui \cite{danrui2021honeycomb}. Structural details of the four models
are summarized in the Supplementary Information. For \rucl, the R$\bar3$ results are shown in Supplementary Information due to very similar results to the C2/m ones.

\begin{figure}
    \centering
    \includegraphics[width=0.9\linewidth]{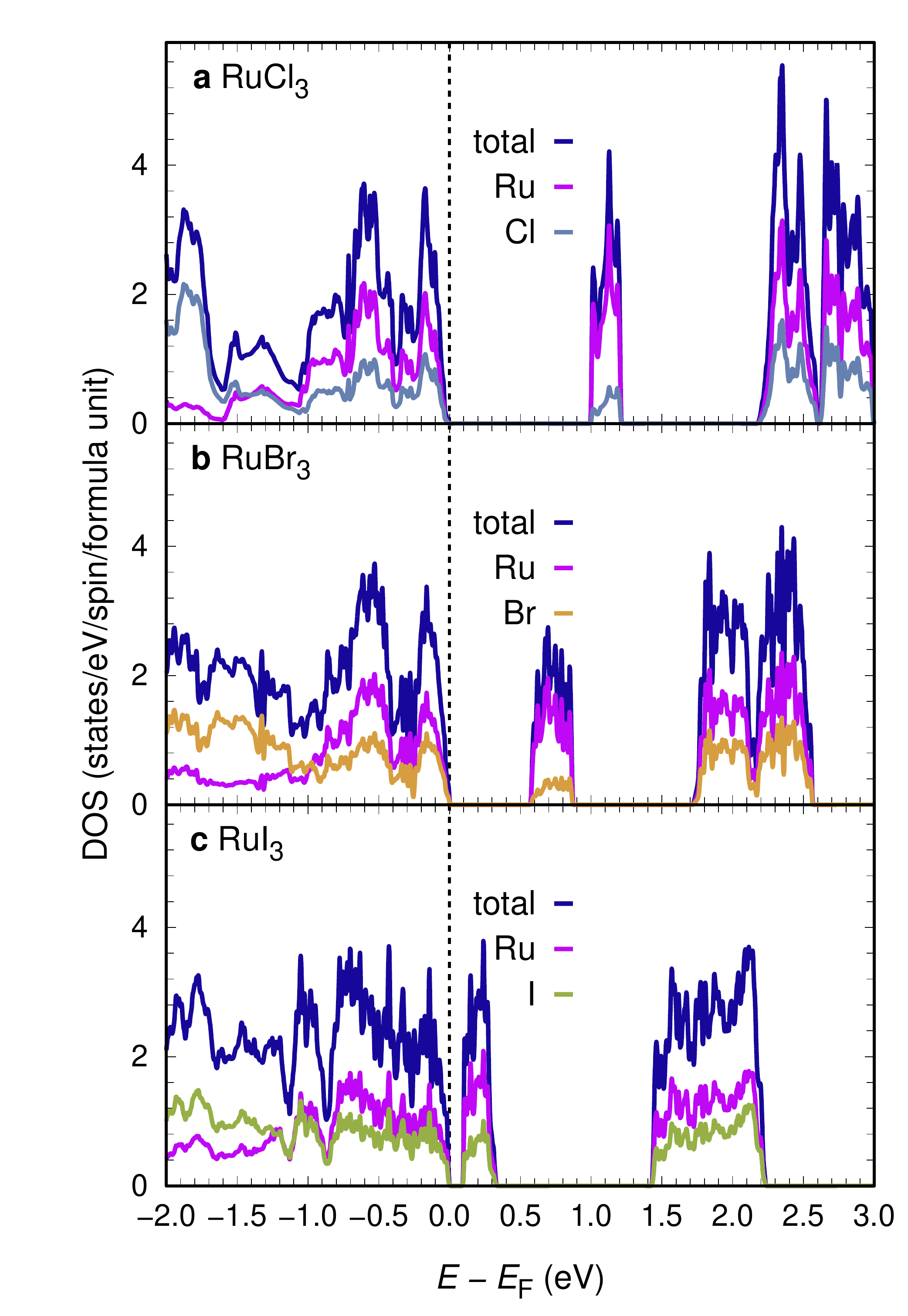}
    \caption{ \textbf{Density of states for \rux (X=Cl, Br, I)} 
    Density of states (DOS) for the experimental structures of \rucl,\rubr\ and \rui, obtained from GGA+SO+U calculations with Wien2k, considering antiferromagnetic zigzag magnetic configurations. For \rucl we employed $U_{\rm eff} = 2.7$\,eV, for \rubr $U_{\rm eff} = 2.1$\,eV and for \rui $U_{\rm eff} = 1.4$\,eV. Shown is also the contribution of Ru and halogen states to the DOS.}
    \label{fig:dos_expstruct}
\end{figure}

\begin{figure*}
    \centering
    \includegraphics[width=\textwidth]{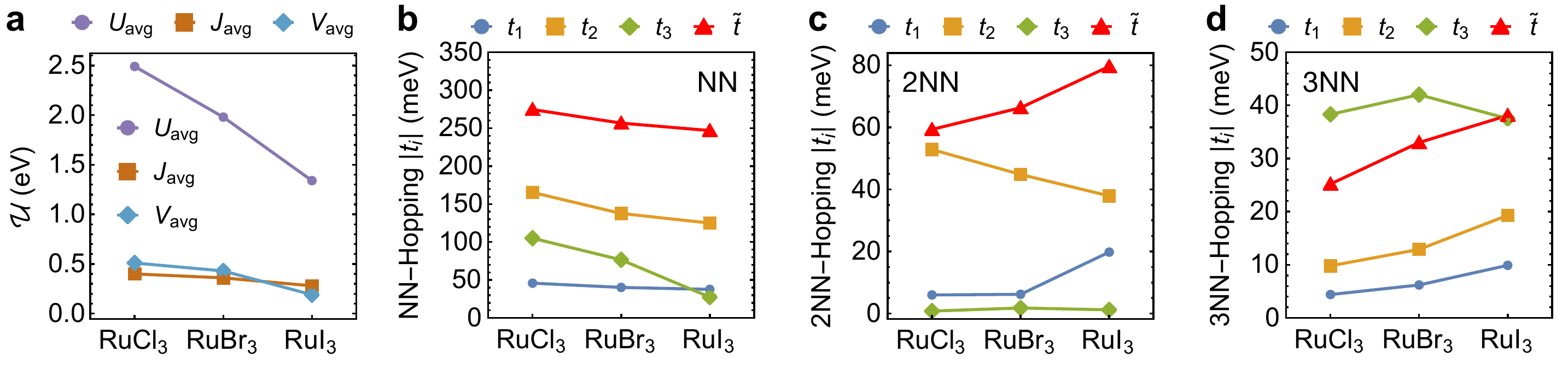}
    \caption{\textbf{\textit{Ab-initio}-computed multi-orbital Hubbard model parameters across the \rux family.}  \textbf{a}~cRPA results for the orbitally-averaged on-site Hubbard interaction ($U_{\rm avg}$), Hund's coupling ($J_{\rm avg}$), and the nearest-neighbour $V_{\rm avg}$ coupling.
\textbf{b}~Absolute magnitude of hopping parameters $t_1=t_{(yz,yz)}$, $t_2=t_{(xz,yz)}$, $t_3=t_{(xy,xy)}$, and $\tilde{t}=t_{(xy,z^2)}$ on nearest-neighbor (NN), second-neighbor (2NN) and third-neighbor (3NN) Z-bonds. 
    }
    \label{fig:model_parameters}
\end{figure*}

\Cref{fig:dos_expstruct} shows the relativistic density of states (DOS) 
obtained within GGA+SOC+U as implemented in Wien2k, 
where a zigzag magnetic configuration with magnetic moments polarized perpendicular to the $ab$ plane was considered.  
For the choice of $U_{\rm eff} =U - J$ 
we take as a reference the {\it ab initio} estimates for the
orbitally-averaged Hubbard  on-site ($U_{\rm avg}$) and Hund's coupling ($J_{\rm avg}$) 
as obtained from constrained random-phase approximation (cRPA) calculations (see "Methods" section for calculation details). In contrast to previous cRPA estimates for \rucl 
\cite{eichstaedt2019deriving}, our estimates incorporate all five $d$ orbitals and
extend to the complete Ru-based trihalide family.
     As shown in \cref{fig:model_parameters}a, the effective Hubbard interaction parameters decrease with increasing ligand atomic number from Cl to I, 
     which can be attributed to the more delocalized nature of the Ru $d$ orbitals in \rui compared to \rucl when hybridizing with I instead of Cl.

For \rucl a $U_{\rm eff}=2.7$\,eV yields both the fundamental 
and direct gap to be  $\approx$ 1\,eV (\Cref{fig:dos_expstruct}a)
in agreement with the reported optical gap, apart from the presence of multiplets at 
200\,meV~\cite{sandilands2016}.  We systematically reduced $U_{\rm eff}$ to 2.1\,eV for \rubr and 1.4\,eV for \rui following the trend given by the cRPA results.  With these values, \rubr shows a gap of 0.56\,eV 
(\Cref{fig:dos_expstruct}b), while \rui shows a small gap of 0.1\,eV (\Cref{fig:dos_expstruct}c). The gap closes in \rui when $U_{\rm eff}$ is further reduced to 1\,eV.
These results indicate a spin-orbit assisted Mott insulating state in disorder-free   \rui samples, which is on the verge of a metal-insulator transition.
Possibly, as discussed above, disorder in the experimental samples could act as effective pressure, and bring the samples closer or over the Mott transition as seen in other Mott insulators~\cite{gati2018effects}.
Note that these results hold regardless of the assumed magnetic pattern in the 
calculations.

In order to analyze the magnetic structure of the \rux compounds, we first consider spin-polarized total energy calculations with VASP in the GGA+SOC+U approximation (see also `Methods'). Detailed results are listed in the Supplementary Information. 
For \rucl, the calculated energy of the ferromagnetic state $E_\text{FM}$
is very competitive with the energy of the experimentally observed zigzag ordered state
$E_\text{ZZ}$: $ E_\mathrm{ZZ}-E_\mathrm{FM}\approx 2$\,meV/Ru.
This observation is consistent with the evidence for a metastable ferromagnetic state in \rucl \cite{bachus2020ThermodynamicPerspectiveFieldInduced,suzuki2021proximate}. Correspondingly, in our effective pseudospin model of \rucl discussed below, 
classically, the energy of the ferromagnet is below that of zigzag, and only by including quantum fluctuations the zigzag ground state is recovered (as in, e.g., Ref.~\cite{suzuki2021proximate}). 
For \rubr we find an energy minimum for the zigzag ordering in agreement with the experiment.
 Interestingly, for \rui N\'eel and zigzag orders
 are energetically almost degenerate
 $ E_\text{N\'eel}-E_\mathrm{ZZ}\approx 1$\,meV/Ru,
 with the rest of 
 magnetic orders we scanned
 being energetically rather close. 
 All orders show very small and varying magnetic moments for Ru. These results hint to a magnetic frustration. %

We proceed with the derivation of magnetic exchange models.  In the first place,
the magnetic Hamiltonian of \rubr has been suggested to be closer to the pure Kitaev limit than in \rucl \cite{imai2021magnetism}, and, secondly,
with our proposed scenario of a Mott insulating state for \rui, the question of its magnetic properties is open. To investigate these issues from first principles, we derive via the \textit{ab-initio} projED method \cite{riedl2019abinitio} the pseudospin models  $\mathcal{H}_{\rm eff}=\sum_{ij} \mathbf{S}_i \cdot \mathbb{J}_{ij} \cdot \mathbf{S}_j$  of the three \rux compounds. 
Here, $\mathbf S$ stands for the relativistic pseudospin $j_\mathrm{eff}=1/2$ moment \cite{jackeli2009mott}.

In the conventional parametrization of Kitaev materials, the exchange matrix $\mathbb{J}_{ij}$ in R$\bar 3$ symmetry on a nearest-neighbor Z$_1$-bond (defined in \cref{fig:structure})
 follows the form
 \begin{align}
    \mathbb{J}_{ij} = \left(\begin{array}{ccc} 
    J_1+\nu_1 & \Gamma_1 & \Gamma^\prime_1 + \eta_1 \\ 
    \Gamma_1 & J_1-\nu_1 & \Gamma_1^\prime - \eta_1 \\ 
    \Gamma^\prime_1 + \eta_1 & \Gamma_1^\prime - \eta_1 & J_1+K_1 \end{array} \right),
    \label{eq:exchange_matrix}
\end{align}
with the isotropic Heisenberg exchange $J_1$, the bond-dependent anisotropic Kitaev exchange $K_1$, the bond-dependent off-diagonal exchange terms $\Gamma_1$ and $\Gamma^\prime_1$ and correction terms $\eta_1$ and $\nu_1$. The latter correction terms are found to be small in our calculated Hamiltonians, and are neglected in what follows.  The exchange matrices on X- and Y-bonds follow by respective $C_3$ rotations about the out-of-plane axis ($[111]$ in pseudospin coordinates).  
Analogously follow the definitions for second and third neighbor exchange terms (or see, e.g., Ref.~\onlinecite{winter2016challenges}).

\begin{figure*}
    \centering
    \includegraphics[width=\textwidth]{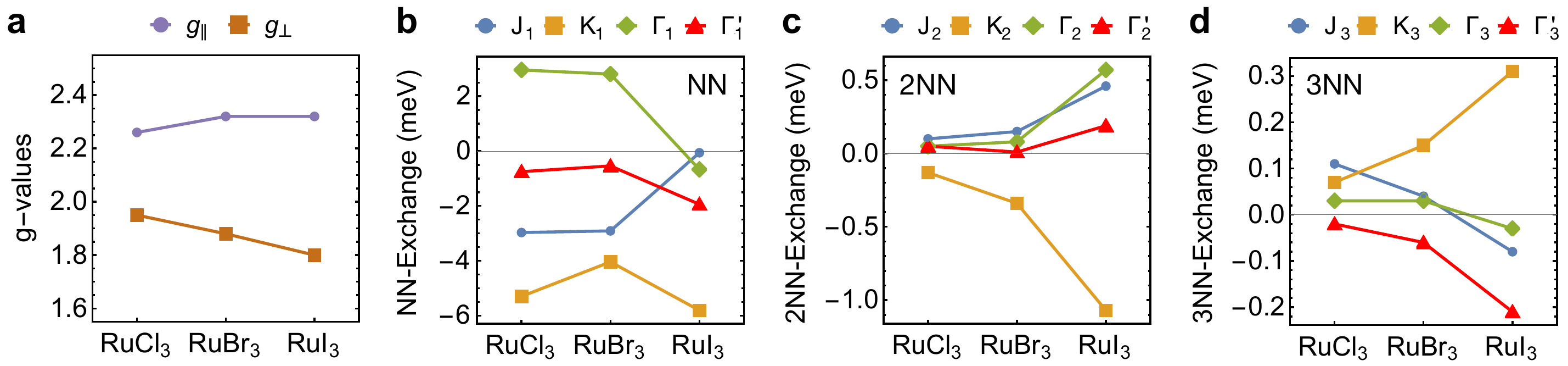}
    \caption{\textbf{\textit{Ab-initio} computed pseudospin models across the \rux family.}  \textbf{a}~Quantum chemistry results for local gyromagnetic \textit{g}-tensor components $g_\parallel$ (in-plane) and $g_\perp$ (out-of-plane). \textbf{b-d}~projED results for the magnetic exchange couplings on nearest-neighbor~(NN), second-neighbor~(2NN) and third-neighbor~(3NN) bonds. 
    Tabular form of all values is given in the Supplementary Information. %
    }
    \label{fig:exchange_parameters}
\end{figure*}

Using $U_{\rm avg}$ and $J_{\rm avg}$ from cRPA (\cref{fig:model_parameters}a),
the complex hopping parameters extracted from full-relativistic DFT (magnitudes shown in \cref{fig:model_parameters}b,c,d)
and the \mbox{projED} method, we extracted the exchange constants shown
in \cref{fig:exchange_parameters}b,c,d.

Evaluating the magnetic interactions of the complete \rux family, we find a nearest-neighbor ferromagnetic Kitaev interaction $K_1$ to be the dominant 
in all three compounds. Additionally, a subdominant ferromagnetic nearest-neighbor Heisenberg exchange $J_1$ is present, which is, however, almost vanishing for the iodine case. The symmetric off-diagonal $\Gamma_1$ interaction is of similar magnitude as $J_1$, changing sign going from Cl and Br to I. $\Gamma^\prime_1$, often neglected in the \rucl analysis, may become rather important,
particularly for \rui. Further-neighbor interactions
are generally smaller than their nearest-neighbor counterparts
for all three systems, but increase for larger ligand atomic number and may play, especially in \rui, an important role.

That the anisotropic interactions do not monotonically increase with stronger spin-orbit coupling of halogen elements can be related to the SOC source. In the original Jackeli-Khaliullin mechanism \cite{jackeli2009mott}, the heavy magnetic ions are solely responsible for SOC effects, which can be well described within the SOC atomic limit. In the case of \rubr and \rui, however, ligand SOC starts to play an important role. To evaluate the interplay of these two SOC sources, we extracted {\it ab initio} values for the \rux materials and compared them to the SOC atomic limit (see Supplementary Information). We find that in these compounds SOC effects from magnetic ions and ligands do not enhance each other, but do compete. This leads to the observed inhomogeneous behavior of the magnetic anisotropic terms in \cref{fig:exchange_parameters} as a function of ligand atomic number.
Another consequence of this breakdown of the SOC atomic limit is that the established analytic perturbation theory expressions~\cite{jackeli2009mott,rau2014generic,winter2016challenges} become unjustified in Kitaev materials where SOC arises from
both the metal and the ligand elements. \rubr and \rui are therefore cases where more general approaches, like ours, are indispensable. Another approach would be perturbation theory taking into account ligand orbitals, as recently derived for the $S=3/2$ material CrI$_3$ \cite{stavropoulos2021magnetic}.

 Along the halogen series Cl-Br-I we observe a decrease for nearest-neighbor couplings (\cref{fig:exchange_parameters}b) and an overall increase in magnitude for second and third neighbors (\cref{fig:exchange_parameters}c,d). This can be understood by consideration of the ligand-metal ($p$-$d$) hybridization. 
We quantify the hybridization strength by integrating the DFT(GGA) density of states (DOS) with Ru $4d$ orbital character in the energy window dominated by the ligand $p$ orbitals (between -7\,eV and \mbox{-1.05\,eV}). 
In spite of respective larger Ru-Ru distances, %
this can be related to the magnetic exchange by consideration of the \textit{ab initio} hopping parameters between Wannier $d$ orbitals. %
As also pointed out in Ref.~\onlinecite{kim2021spin}, in spite of the stronger hybridization the nearest-neighbor hopping parameters are reduced for heavier ligands,
illustrated in \cref{fig:model_parameters}b.
This is reflected in the magnetic exchange parameters
(\cref{fig:exchange_parameters}b) in an overall reduced magnitude in the nearest-neighbor parameters. 
In contrast, the second and third neighbors %
show a very different dependence on the halogen element. From the dominant further-neighbor hoppings (\cref{fig:model_parameters}c,d), the hoppings show an overall tendency to increase, with few exceptions. 
Certain further-neighbor magnetic exchange parameters, depending on their relation to the individual hopping parameters, become therefore increasingly important for \rubr and especially for the magnetic properties of \rui.

Finally, we also computed the gyromagnetic $g$-tensor for the \rux family from first principles, in order to relate the pseudospin $\mathbf S$ of the effective Hamiltonian to the magnetic moment $\mathbf M = \mu_\text{B} \mathbb G \cdot \mathbf S$.  
The $g$-tensor can be approximately characterized by two components, the value parallel to the honeycomb plane, 
$g_{\parallel}$, and the one perpendicular to it, $g_{\perp}$, which are shown in \cref{fig:exchange_parameters}a. We consistently find $g_{\parallel}>g_{\perp}$ for the whole family, promoting a stronger Zeeman term for in-plane fields.

We now discuss the ramifications of the derived magnetic
models for the magnetism in these materials. 

\label{link123} For  \rucl, we can compare our result to a vast available literature of models that have been shown to reproduce various experimental observations. %
Indeed, the model presented here in \cref{fig:exchange_parameters}, derived completely from first principles without adjustments or external parameters, is remarkably close to some well-benchmarked recent models \cite{winter2017breakdown,suzuki2021proximate,kaib2021magnetoelastic}, and is therefore expected to also describe the material quite well. 
As we apply the same \textit{ab-initio} setup for the new members of the \rux family, we expect our models to be reliable for them too.

 \begin{table}
\centering%
\begin{ruledtabular}
\begin{tabular}{c|cccc}
 RuX$_3$
      & RuCl$_3$ & RuBr$_3$ & RuI$_3$   \\
      \hline 
  $\Theta_\text{CW}^\text{avg}$ & +39.1\,K&+35.6\,{K}&+15.4\,{K}\\ 
  \hline 
  \text{GS} & \text{Zigzag} & \text{Zigzag} & \text{QSL?} 
  \\ $\phi_{\mathbf M}$ & $90^\circ$ & $90^\circ$ & 
  \\ $\theta_{\mathbf M}$ & $34.4^\circ$ & $32.4^\circ$ & 
  \\
\end{tabular}
\end{ruledtabular}
\caption{
\textbf{Properties of derived pseudospin models.}
$\Theta_\text{CW}^\text{avg}$ is the powder-averaged Weiss temperature of each model. `GS' refers to the ground state computed by exact diagonalization, and the angles of the magnetic moment $\phi_\mathbf{M},\theta_{\mathbf M}$ are defined according to \cref{fig:structure}b. } 
\label{tab:summary_model_properties}
\end{table}

The direction-averaged Weiss constant ($\Theta_\text{CW}^\text{avg}$ in \cref{tab:summary_model_properties}) 
is predicted to be positive across the \rux family, characteristic of %
{ferromagnetic} %
exchange interactions.
This is in line with our analysis of the experimental magnetic susceptibilities of \rucl and \rubr above (\cref{fig:susceptibility}).
While in \rucl and \rubr a large FM contribution to the Weiss constant comes from a significant FM nearest-neighbor Heisenberg interaction $J_1$, this interaction nearly vanishes for \rui (\cref{fig:exchange_parameters}b), leading to a smaller Weiss constant. Furthermore, the small $J_1$ in \rui renders the nearest-neighbor interactions to be extremely anisotropic, with a dominant Kitaev interaction $K_1$. While at first glance this might suggest 
a spin-liquid ground state in \rui, the increased strength of the
further-neighbor interactions in \rui (see, e.g., $J_2,$ $K_2$ in \cref{fig:exchange_parameters}c) also needs to be considered %
\cite{rousochatzakis2015phase}.

\begin{figure*}
\centering
\includegraphics[width=\textwidth]{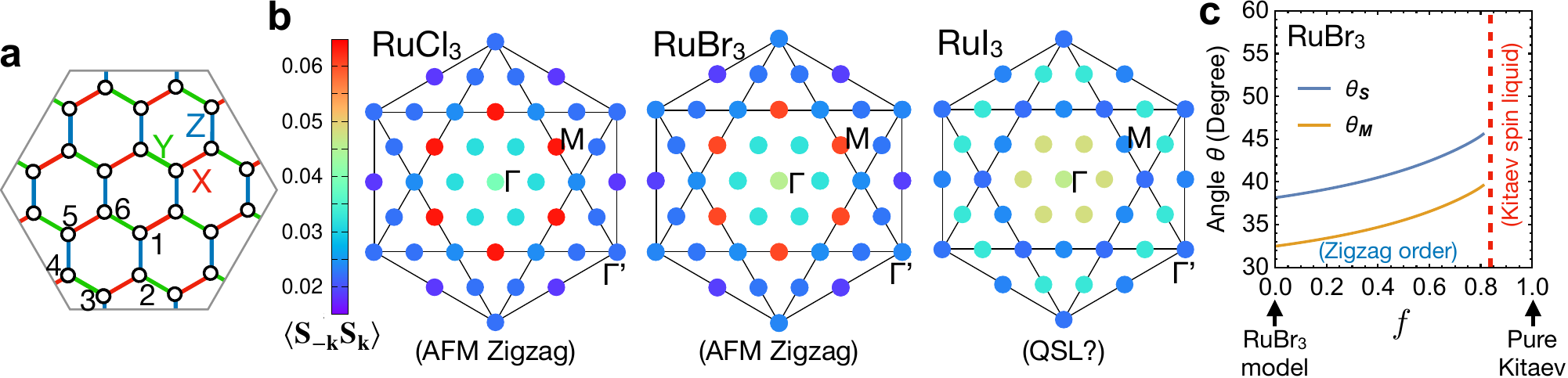}
 \caption{\textbf{Exact diagonalization of \rux pseudospin models.} \textbf{a}~Employed periodic cluster. Labeled sites $1,\dots,6$ define the Kitaev plaquette operator $W_p=2^6 S_1^x S_2^y S_3^z S_4^x S_5^y S_6^z$. \textbf{b}~Static spin structure factor in reciprocal space. Inner (outer) hexagon mark the edge of the first (third) Brilluoin Zone. High-symmetry $k$-points $\Gamma,\mathrm M, \Gamma'$ are labelled. Color scale is the same for all three plots. \textbf{c}~Out-of-plane angle $\theta_{\mathbf S}$ ($\theta_{\mathbf M}$) of the pseudospin (magnetic moment) within the zigzag phase when tuning from the \rubr model ($f=0$) towards the pure Kitaev model ($f=1$). Dashed vertical line indicates phase transition to the Kitaev spin liquid, identified by a peak in $-\partial^2 E / \partial f^2$. 
    } 
    \label{fig:EDresults}
\end{figure*}

To find the magnetic ground state properties, we perform 
exact diagonalization (ED) calculations of the derived $j_\text{eff}=1/2$ models on the 24-site cluster shown in \cref{fig:EDresults}a. In  \cref{tab:summary_model_properties} and \cref{fig:EDresults}b we summarize the encountered ground states, i.e.\ zigzag for \rucl and \rubr, and possibly a quantum spin liquid (QSL) in \rui. This is discussed in detail below.

For \rucl, the model in \cref{fig:exchange_parameters} (as well as the R$\bar3$ model discussed in the Supplementary Information) yields zigzag AFM order, identifiable by a maximum at $\mathbf k=\mathrm M$ in the static spin structure factor, shown in \cref{fig:EDresults}b. 
The computed ordered magnetic moment direction (see `Methods' section), parametrized by $\theta$ and $\phi$ in \cref{tab:summary_model_properties} (compare \cref{fig:structure}b), is found to be tilted by
$\theta_{\mathbf M}\approx
34^\circ$ out of the plane, in excellent agreement with 
the recent experiment, where $\theta_{\mathbf M}=32\pm 3^\circ$ 
\cite{sears2020ferromagnetic} was reported. Interestingly, on the classical level, the ferromagnetic state is lower in energy than the zigzag state, meaning that the latter only becomes the ground state through quantum fluctuations, as discussed also in Ref.~\cite{suzuki2021proximate}. 
 
We will now focus on the recently synthesized compounds, starting with \rubr. The static spin structure factor for the \rubr model of \cref{fig:exchange_parameters} is shown in \cref{fig:EDresults}b, indicating also a zigzag AFM  order ($\mathbf k=\mathrm M$ and $C_6$-rotated vectors), in agreement with experiment \cite{imai2021magnetism}. However, the calculated tilt angle of the magnetic moment, $\theta_{\mathbf M}=32^\circ$,  %
is more in line with \rucl than with the reported measured $\theta_{\mathbf M}=64^\circ$ of \rubr \cite{imai2021magnetism}. 
The authors of Ref.~\onlinecite{imai2021magnetism} argued that this anomalously large tilt angle indicates an exceptionally strong relative Kitaev coupling,
\textit{i.e.}, larger $|K_1/J_1|$ and $|K_1/\Gamma_1|$ compared to \rucl. 
To investigate to what extent a closer proximity to the pure Kitaev model could produce such high tilt angles, we take our \rubr Hamiltonian of \cref{fig:exchange_parameters} as a starting point and tune towards the pure Kitaev model, where $K_1$ is the only non-zero
coupling. This is done by multiplying every exchange coupling except $K_1$ by $(1-f)$ and sweeping $f$ from 0 to 1. As shown in \cref{fig:EDresults}c, the moment indeed rotates further away from the honeycomb plane upon moving towards the pure Kitaev model, however even right before the transition to 
the Kitaev spin liquid (indicated by the vertical dashed line), $\theta_{\mathbf S}$ does not exceed $46^\circ$.  
$\theta_{\mathbf M}=\arccos\left(\frac{\cos\theta_{\mathbf{S}}}{\sqrt{{g_{\parallel}}^2\cos^2\theta_{\mathbf S}+{g_{\perp}}^2\sin^2\theta_{\mathbf S}}}\right)$, which is to be compared to the neutron diffraction experiment, is even smaller due to the anisotropy $g_{\rm \parallel}>g_{\rm \perp}$ in our calculated $g$-tensor (\cref{fig:exchange_parameters}a). A reconciliation with the reported $\theta_{\mathbf M}=64^\circ$ would therefore require quite drastic changes to the $g$-tensor anisotropy and/or the exchange parameters. While in the whole $J_1$-$K_1$-$\Gamma_1$ parameter space with $\Gamma_1>0$, no angles of $\theta_{\mathbf S}$ beyond $\sim40^\circ$ are expected in the zigzag phase \cite{rusnacko2019KitaevlikeHoneycombMagnetsa}, significant \textit{negative} $\Gamma_1<0$ can in principle lead to $\theta_{\mathbf S}$ beyond 60$^\circ$ \cite{chaloupka2016MagneticAnisotropyKitaev}.   However such terms seem incompatible with the \textit{ab-initio} results and would likely need strong distortions from the present considered \rubr crystal structure to be realized.

More distinct from the other two compounds are our results for \rui. As discussed above, in our GGA+SOC+U calculations we find a very flat energy landscape of competitive magnetic configurations, indicative of strong magnetic frustration. Fittingly, the ground state from exact diagonalization of the present exchange model %
does not show a dominant ordering wave vector in the spin structure factor, see \cref{fig:EDresults}b. 
Although this is a signature generally associated with quantum spin liquid (QSL) states, we note that
in the present model, the Kitaev $\mathbb Z_2$ flux operator yields $\langle W_p\rangle=2^6 \langle S_1^x S_2^y S_3^z S_4^x S_5^y S_6^z\rangle \approx 0.29$ (site indices refer to \cref{fig:EDresults}a). 
While this is clearly elevated compared to classical collinear states, where $\langle W_p \rangle$ is restricted to $|\langle W_p\rangle|\le \frac{1}{27}<0.04$%
, it is still significantly below the value of the pure unperturbed Kitaev spin liquid, where $\langle W_p\rangle=1$  \cite{kitaev2006anyons}. Hence, if the ground state constitutes a QSL state, it is presumably not the $\mathbb Z_2$ Kitaev spin liquid. 
The precise nature of the encountered magnetically disordered state might be interesting for future studies. It appears to be stabilized by the further-neighbor interactions, as we find a clear ferromagnetic ground state when omitting the second- and third-neighbor interactions in the present model.  %
While a QSL scenario for our full \rui\ model is compelling, we note that finite-size effects in our calculation could play a role. In particular, the finite-size cluster could be incompatible with the supposed correct ordering wave vector of the model, e.g.~in case of an incommensurate ordering vector.

\subsection{Conclusions and outlook}
To summarize, we have presented a comparative analysis of the electronic and magnetic properties of the
Ru-based trihalide family, including the recently synthesized \rubr and 
\rui, by combining state-of-the-art {\it ab initio} microscopic modelling 
with analysis of reported resistivity, specific heat and magnetic susceptibility
data. The evolution of the magnetic order and Mott-Hubbard correlations
along the halogen series, as well as possible role of disorder, have been a central part of our study.
We conclude that:

\begin{enumerate}

\item All three ideal compounds are spin-orbit-assisted Mott insulators, 
but their fundamental gap decreases with higher
ligand atomic number, Cl$\rightarrow$Br$\rightarrow$I, with \rui coming rather close to a metal-insulator transition.
    \item
From DFT total-energy calculations, in ideal, pristine crystals the zigzag magnetic
order is even more stable in \rubr than in \rucl, while \rui shows significant magnetic frustration.
 Our {\it ab-initio} extracted low-energy models predict \rui to feature either an incommensurate magnetic ordered state or a quantum spin liquid, which, interestingly,
is possibly of a different kind to the $\mathbb Z_2$ Kitaev spin liquid.  

\item A number of reported experimental observations seem to be
adversely affected by the sample quality, in particular by dirty grain boundaries. In fact, 
most of the observations in \rui can be reconciled with theory by assuming insulating grains surrounded by (bad) metallic boundaries. 
The experimental evidence is consistent with a `dirty' insulator, or a bad metal. Disorder would favor either of these. 

\item In all three systems the dominant nearest-neighbor interaction is FM Kitaev $K_1$, with a subdominant FM Heisenberg Interaction $J_1$, that nearly vanishes for \rui.
We observe a non-monotonous behavior of the magnetic anisotropic terms as a function of ligand atomic number
that we trace back to a competition of the SOC effects from magnetic ions and ligands. %

\item 
\rubr has predominantly {ferromagnetic} interactions, in contrast to what is suggested by standard Curie-Weiss analysis \cite{imai2021magnetism}. 
Such interactions are consistent with the experimental susceptibility when taking into account high-temperature SOC effects. Our {\it ab-initio} magnetic model predicts zigzag order in agreement with experiment, with a tilting angle
of $\theta_{\mathbf M}=32^\circ$ for the magnetic moments, similar to \rucl,
but in contradiction to the reported $\theta_{\mathbf M}=64^\circ$ \cite{imai2021magnetism}. We showed that such a large angle cannot be simply explained by proximity to the pure Kitaev model, but would require quite drastic changes to the exchange parameters, such as
sizeable negative $\Gamma_1<0$. Those would necessitate strong distortions on the reported \rubr crystal structures. 
\end{enumerate}

Answering the question posed in the title, our results and analysis strongly suggest that the ideal \rucl, \rubr and \rui compounds constitute a family of three Mott-insulating \emph{siblings}. 
The challenging task
of getting better samples will hopefully help resolve the open issues.

\section{Methods}
\subsection{Modified Curie-Weiss fit of \rubr \label{sec:appendix_suscept}}
We fit the experimental average susceptibility of Ref.~\onlinecite{imai2021magnetism} with four fitting parameters $\chi_0^{\perp}, \chi_0^{\parallel}, \Theta^{\perp}, \Theta^{\parallel}$ %
using the modified Curie-Weiss formula
\begin{align} 
\chi^{\rm avg}(T) \approx &  \frac{2}{3}\left(\chi_0^{\parallel}  + \ \frac{C^{\parallel}(T)}{T-\Theta^{\parallel}}\right) + \frac{1}{3}\left( \chi_0^{\perp}  + \ \frac{C^{\perp}(T)}{T-\Theta^{\perp}}\right) ,
\label{eq:sus}
\end{align}
where $\Theta^{\parallel}$, $\Theta^{\perp}$ are the Weiss constants and $C^\alpha(T) \propto [\mu_{\rm eff}^\alpha(T,\Delta)]^2$ is determined through $\Delta$ as described in Ref.~\onlinecite{li2021ModifiedCurieWeissLaw}. 
Superscripts $\parallel$ and $\perp$ indicate the in- and out-of-honeycomb-plane direction respectively. The susceptibility is fitted over the temperature range 150 -- 300\,K and SOC strength $\lambda=0.15\,$eV is taken. %

\subsection{DFT calculations}

To make sure that the calculated features within density functional theory are robust with respect to the choice of the basis set, we have tested the results using two different methods:
the projector augmented wave method~\cite{blochl1994,kresse1999} as implemented in
the VASP code~\cite{kresse1993,kresse1996}, and the full potential linearized augmented plane-wave (LAPW) basis as implemented in Wien2k~\cite{Wien2k}. Throughout the paper we have
used the Generalized Gradient Approximation  (GGA~\cite{perdew1996}) to the
exchange-correlation functional. Hubbard correlation effects  were included on a
mean field level in the rotationally invariant implementation
of the GGA+U method~\cite{Anisimov1993}. 
All calculations included  spin-orbit  coupling  (SOC)  effects.   
   For VASP we used the Ru\_pv pseudopotential, treating Ru $p$ states as valence, and the standard pseudopotentials for the halogens. The $\Gamma$-centered $8\times8\times8$ mesh in the nonmagnetic rhombohedral Brillouin zone was used, or the correspondingly scaled
meshes for other structures. The energy cut-off was 350\,eV, and the energy convergence criterion 1$\times10^{-08}$\,eV. For each type of magnetic order a number of collinear 
starting configurations with randomly selected N\'eel vectors were used, and the
lowest-energy result was selected as the ground state. Individual results can be found
in the Supplementary Information. 
For Wien2k we chose the plane-wave cutoff $K_{\rm max}$ corresponding to  RK$_{\rm max} = 8$ and a {\bf k}  mesh of $8\times8\times2$  for the R$\bar{3}$ structure in the 
hexagonal Brillouin zone and $8\times4\times6$ in the first Brillouin zone of the conventional unit cell for the $C/2m$ structure. The density of states are calculated using a {\bf k} mesh of $12\times12\times3$ for the R$\bar{3}$ structure and $12\times6\times9$ for the $C/2m$ structure.
The zigzag configurations are constructed using a conventional cell of the $C/2m$ structure for \rucl while a $1 \times 2 \times 1$ supercell of the R$\bar{3}$ structures for \rubr and \rucl. %

\subsection{%
cRPA calculations}
In order to obtain {\it ab-initio} estimates for the effective Coulomb interaction for the Ru-trihalide family, we employed the constrained random-phase approximation (cRPA)~\cite{Aryasetiawan2004,Aryasetiawan2006}, as implemented in the FHI-gap code~\cite{fhigap}, based on the Wien2K electronic structure.
The low-energy limit of the screened interaction was projected on the five Ru $d$ orbitals, where screening processes in the same window were excluded. Convergence with respect to the discretization of the Brillouin zone and energy cutoff was ensured.

\subsection*{DFT-based derivation of magnetic models}

To derive bilinear exchange parameters for each material, we employed the projED method \cite{riedl2019abinitio}, which consists of two steps. 
First, complex {\it ab-initio} hopping parameters between the ruthenium ions are estimated with projective Wannier functions \cite{eschrig2009} applied on full relativistic FPLO~\cite{koepernik1999} calculations on a $12 \times 12 \times 12$ $\mathbf{k}$ mesh. This allows to construct an effective electronic model $\mathcal{H}_{\rm tot} = \mathcal{H}_{\rm hop}+ \mathcal{H}_{\rm U}$, where the complex {\it ab-initio} hopping parameters enter the kinetic term $\mathcal{H}_{\rm hop} = \sum_{ij\alpha \beta} \sum_{\sigma \sigma^\prime} t_{i\alpha,j \beta}^{\sigma \sigma^\prime}\, c_{i\alpha\sigma}^\dagger c_{j\beta \sigma^\prime}$ and the cRPA effective Coulomb interaction parameters enter the two-particle term $\mathcal{H}_{\rm U} = \sum_{i \alpha \beta \gamma \delta} \sum_{\sigma \sigma^\prime} U_{i \alpha \beta \gamma \delta}^{\sigma \sigma^\prime}\, c_{i \alpha \sigma}^\dagger c_{i \beta \sigma^\prime}^\dagger c_{i \delta \sigma^\prime} c_{i \gamma \sigma}$.
Second, the effective spin Hamiltonian $\mathcal{H}_{\rm eff}$ is extracted from the electronic model via exact diagonalization (ED) and projection of the resulting energy spectrum onto the low-energy subspace, mapped onto pseudo-spin operator representation in the $j_{\rm eff}$ picture with the projection operator $\mathbb{P}$: $\mathcal{H}_{\rm eff} = \mathbb{P} \mathcal{H}_{\rm tot}\mathbb{P} = \sum_{i j } \mathbf{S}_i\, \mathbb{J}_{ij}\, \mathbf{S}_j$. 

Note that for \rucl the exchange constants slightly differ from previously calculated values by some of the authors~\cite{winter2016challenges,kaib2021magnetoelastic}.
The reason for this lies in the following details of the calculation setup: (i) first principles input parameters $U_{\rm avg}$ and $J_{\rm avg}$ from cRPA in contrast to 
previous choices, (ii) consideration of all five $4d$ ruthenium orbitals with the cost of restriction onto two-site clusters, (iii) SOC effects from both $\mathrm{Ru}^{3+}$ and ligands considered through complex hopping parameters in contrast to the atomic limit, and (iv) consideration of the experimental crystal structure in contrast to relaxed ambient pressure structure as it was done in Ref.~\onlinecite{kaib2021magnetoelastic}.

For the calculation of the gyromagnetic $g$-tensor, we considered [RuX$_6$]$^{3-}$ molecules within the quantum chemistry ORCA 3.03 package~\cite{neese2012orca,neese2005efficient} with the functional TPSSh, basis set def2-TZVP and complete active space for the $d$ orbitals CAS(5,5). 

\subsection*{Exact diagonalization}
Exact diagonalization calculations of the $j_\text{eff}=1/2$ models were performed on the 24-site cluster shown in \cref{fig:EDresults}a. 
To identify possible magnetic ordering, we analyze the static spin structure factor $\sum_{\mu=x,y,z}\langle S^\mu_{-\mathbf k} S^\mu_{\mathbf k} \rangle$. For the ordered moment direction, we compute the eigenvector with maximal eigenvalue of the correlation matrix $(\langle S^\mu_{-\mathbf k}S^\nu_{\mathbf k} \rangle)_{\mu,\nu}$ ($\mu,\nu \in \{x,y,z\}$) at the ordering wave vector $\mathbf k=\mathbf Q$. 
This eigenvector then represents the ordered \textit{pseudospin} direction $\mathbf S$~\cite{chaloupka2016MagneticAnisotropyKitaev},  which relates to the magnetic moment direction $\mathbf M \propto \mathbb G \cdot \mathbf S$, as measured by neutron diffraction, via the anisotropic $g$-tensor $\mathbb G$.%



\section{Acknowledgments}
We thank Stephen M.\ Winter, Robert J.\ Cava, Yoshinori Imai, and Elena Gati for discussions and Yoshinori Imai for sharing the structural information of \rubr with us before publication. %
R.V., A.R., K.R.\ and D.A.S.K.\ acknowlegde support by the Deutsche Forschungsgemeinschaft (DFG, German Research Foundation) for funding through Project No.\ 411289067 (VA117/15-1) and TRR 288 --- 422213477 (project A05). Y.L. acknowledges support by National Natural Science Foundation of China (Grant No.\ 12004296) and China Postdoctoral Science Foundation (Grant No.\ 2019M660249). I.I.M. acknowledges support from the U.S. Department of Energy through the grant \#DE-SC0021089. 
R.V. and I.I.M. thank the Wilhelm und Else Heraeus Stiftung for financial support.

\section{Competing interests}

 We declare no competing interests. 

 \section{Author contributions}

 R.V.~conceived and supervised the project. Density functional theory calculations were performed by K.R., A.R., Y.L., I.I.M., cRPA calculations by S.B., projED calculations by K.R., and calculations on magnetic models by D.A.S.K. All authors contributed to the manuscript.  

\section{Supplementary information}
The online version contains supplementary material available at ... 
\nocite{Mazin2012,Foyevtsova2013,johnson2015,widmann2019thermodynamic,cao2016,park2016,mu2021RoleThirdDimension,sears2020ferromagnetic,imai2021magnetism,danrui2021honeycomb,nawa2021StronglyElectroncorrelatedSemimetal,Mazin2012,Foyevtsova2013,jackeli2009mott}

\end{document}